# MOTION OF THE MASS CENTERS IN A SCALAR THEORY OF GRAVITATION: THE POINT-PARTICLE LIMIT

MAYEUL ARMINJON

*Laboratoire "Sols, Solides, Structures"*
*[CNRS / Université Joseph Fourier / Institut National Polytechnique de Grenoble]*
*B.P. 53, 38041 Grenoble cedex 9, France.*
*E-mail: arminjon@hmg.inpg.fr*

*Abstract.* We investigate the point-particle limit of the equations of motion valid for a system of extended bodies in a scalar alternative theory of gravitation: the size of one of the bodies being a small parameter $\xi$, we calculate the limit, as $\xi$ tends towards 0, of the post-Newtonian (PN) acceleration of this small body. We use the asymptotic scheme of PN approximation, that expands all fields. We find that the PN acceleration **A** of the small body keeps a structure-dependent part at this limit. In particular, if the only massive body is static and spherical, then **A** differs from the PN acceleration of a test particle in a Schwarzschild field only by this structure-dependent part. The presence of the latter is due to the fact that the PN metric depends on the first spatial derivatives of the Newtonian potential. Since just the same form of PN metric is valid for the standard form of Schwarzschild's solution, the acceleration of a small body might keep a structure-dependent part at the point limit in general relativity also, depending on the gauge. The magnitude of the structure-dependent acceleration is already challenging on Earth. For the Pioneer spacecrafts, it is likely to discard the current version of the scalar theory. A modified version has been outlined in a quoted reference.

## 1. INTRODUCTION

It seems natural that a body of a sufficiently small mass should have a negligible effect on the motion of much more massive bodies. It may also seem natural that its own motion should be determined by its initial conditions and by the trajectories of the much more massive bodies, without the internal structure of the small body having any effect. In a nonlinear field theory, however, it is not immediate to check either of the two foregoing statements. In the theory of gravitation, it is usual to introduce the notion of a test particle (a point particle that precisely follows the foregoing rules), but it is difficult to find in the literature a link between the equations for test particles and those for extended bodies. Fock [1, §81] showed that, *"in the limit* [where the mass of one of the bodies vanishes,] *the integrals of motion in the* [post-Newtonian] *two-body problem* [of general relativity (GR) in harmonic coordinates] *do indeed go over into the corresponding integrals of the equations of the geodesic which determine the motion of an infinitesimal mass in the field of a finite mass.*" However, these two integrals of motion are not enough to determine the motion: one needs also to have the expression of the metric at the world line of the (center of mass of the) small body [Eq. (58.31) in Ref. 1]; this expression is certainly influenced by the presence of that very body, however small be its mass; moreover, Fock's result is limited to the two-body problem for non-rotating, spherically symmetrical rigid bodies. Furthermore, besides the ``standard'' post-Newtonian approximation (PNA), developed by Fock [1] and by Chandrasekhar [2], there is now another, ``asymptotic'' scheme of PNA, which has been initiated for general relativity by Futamase & Schutz [3]; their approach consists in defining a one-parameter family of solutions to the equations of a perfect fluid in GR (in the harmonic gauge), by defining a family of initial conditions. Moreover, Rendall [4] has studied the asymptotic behaviour of *a priori* given families of solutions to the equations of a perfect fluid in GR (in various gauges), which are relevant to the PNA, and his results pertain in fact to the asymptotic PNA. An ``asymptotic'' scheme of PNA, which turns out to be rather close to the scheme proposed by Futamase & Schutz, has also been (independently) proposed [5-6] for an alternative, scalar theory of gravitation, and has then been rather extensively developed [7-10] for this same theory. (See Ref. 11 for a summary of that theory and a proof that it predicts the same effects on light rays and a similar radiative energy loss as does GR. See Ref. 12 for a review of the development of ``asymptotic'' PN celestial mechanics in that theory and its test.) The ``asymptotic'' scheme occurs naturally from the point of view of the approximation theory, for it is an application of the usual method of asymptotic expansion for a system of partial differential equations (whence the name that we use for this scheme).

The application of the asymptotic PNA scheme of the scalar theory to get the explicit equations of motion of the mass centers for a system of well-separated rigidly moving bodies has led to the conclusion that *the internal structure of the bodies does influence the motion*, at least for that theory [10]. It is easy to see that this influence does not cancel if one makes the size of all bodies but one tend toward zero (which is a relevant asymptotics in the solar system). Hence it appears advisable to study in detail the point-particle limit of the translational equations of motion of the asymptotic PNA scheme, and this is the aim of the present paper. However, it turns out that the asymptotic expansions used to express, respectively, the ``good separation'' between bodies and the point-particle limit, are hardly compatible [this point will be demonstrated in § (3.1) below]. Therefore, to study the point-particle limit, one has to start from the general form of the PN equations of motion for the mass centers. This general form has been derived in Ref. 7, to which we shall refer, and of which we shall use the notations. The situation suitable to study the point-particle limit will be defined in *Section 2;* the size of just one body will be assumed small: *a priori,* this could be a satellite or a spacecraft, but still it could be a planet. The bulk of the here-reported work was the study of the asymptotic behaviour of the different integrals that appear in the translational equations of motion (*Section 3*). *Section 4* will be devoted to the case that there is only one massive body, assumed static and spherical. In *Section 5*, we shall analyse the reason for the difference between the equation of motion of the small body at the point-particle limit, and the equation of motion for a test particle; we shall also discuss the connection with the weak equivalence principle. The order of magnitude of that difference will be evaluated (*Section 6*) in two different situations: for small bodies on the surface of the Earth, and for the Pioneer spacecrafts, for which an extremely small ``anomalous" acceleration has been found [13-14]. Our *conclusion* makes *Section 7*, which is followed by a necessary technical Appendix.

## 2. DEFINITION OF THE POINT-PARTICLE LIMIT

We envisage a system S of *N* bodies, each of which is made of a barotropic perfect fluid, and we consider the first PNA of the scalar theory [6-8], which is enough for the solar system. We investigate the situation where the first body, (1), is very small, its ``radius" being a small parameter,

(2.1) $$r_1 = \xi \ll 1.$$

(The radius of body (*a*) may be defined by

$$r_a \equiv \tfrac{1}{2} \mathrm{Sup}_{\mathbf{x},\mathbf{y} \in D_a} |\mathbf{x} - \mathbf{y}|,$$

where $D_a(T)$ is the domain occupied by (*a*) at time $T$ in the preferred reference frame of the theory.) Thus, we consider a family of 1PN systems, depending on $\xi$. It would be easy to define this family explicitly by writing the initial conditions, which are independent of $\xi$ apart from the fact that the size of domain $D_1$ is $r_1 = \xi$. Note that the gravitational potentials: the Newtonian potential associated with the zero-order rest-mass density $\rho$, $U = \mathrm{N.P.}[\rho]$, and the 1PN potentials $B = \mathrm{N.P.}[\sigma_1]$ where $\sigma_1$ is the 1PN correction to the density of active mass [7, Eq. (2.24)], and $W$ [7, Eq. (4.14)], depend merely on the matter fields. Since, moreover, the initial conditions for the 1PN corrections to the independent matter fields of pressure and velocity are simply [6]

(2.2) $$p_1(\mathbf{x},T=0) = 0, \quad \mathbf{u}_1(\mathbf{x},T=0) = \mathbf{0},$$

it follows that the initial conditions which define a 1PN system reduce to those for the Newtonian (i.e. 0PN) density $\rho$ and for the Newtonian velocity field **u** (as it is also the case for a Newtonian system). We shall not need explicit initial conditions and shall merely assume that they ensure that, for all bodies including the small body (1), $\rho$ and **u** are of order $\xi^0$ exactly [this is noted ord($\xi^0$)]. Thus, in particular: *as a smaller body is considered, its density remains of the same magnitude.* The Newtonian potential $U$, as well as the 1PN potential $A = B + \partial^2 W/\partial T^2$ [7, Eq. (4.14)], depend linearly on integrals of $\rho$. Hence the assumption that $\rho$ is $O(\xi^0)$ implies that the part of those potentials which is produced by the small body (1) is $O(\xi^3)$, if one restricts the consideration to distances from (1) which are lower-bounded by a number independent of $\xi$. This implies that the Newtonian acceleration [7, Eq. (4.6)] of any *other* body, (*a*) with $a > 1$, as well as the 1PN correction to it [7, Eq. (4.9), involving the integrals (4.8), (4.10) and (4.18) there], both depend on body (1) by $O(\xi^3)$ terms only.[1] This is true provided the small body remains at a distance from the other bodies which is lower-bounded by a number independent of $\xi$ – in physical terms: provided the small body does not approach any of the other ones by a distance of the same order as its small size $\xi$. (Should the body (1) become nearly tangent to one of the ``large'' bodies, say (2), then (1) would influence (2) by $O(\xi)$ terms, for such would be the order of the contribution of (1) to $U_{,i}$.) *This result provides a rigorous basis for ignoring the influence of a small body on the motion of the other bodies in a 1PN gravitationally-bound system.*

The remainder of the paper will thus concentrate on the motion of the mass center of (1), including the influence of the structure of (1) on that motion. More precisely, we shall study the behaviour of the 1PN correction, considered for $a = 1$. Of course, the main contribution to the acceleration of the mass center of (1) is the Newtonian part, but this part is just the barycenter of the Newtonian acceleration due to the external bodies and it is hence unproblematic as regards the point-particle limit: a Taylor expansion gives us

(2.3) $\qquad U^{(a)}_{,i}(\mathbf{x}) = U^{(a)}_{,i}(\mathbf{a}) + (x^j - a^j)\, U^{(a)}_{,ij}(\mathbf{a}) + O(\xi^2) \quad (a = 1,\ \mathbf{x} \in D_1).$

From the definition of the Newtonian mass center $\mathbf{a}$ and the Newtonian equation of motion, we get then:

(2.4) $$\ddot{\mathbf{a}} = \nabla U^{(1)}(\mathbf{a}) + O(\xi^2).$$

Henceforth, $\mathbf{a}$, $\dot{\mathbf{a}}$ and $\ddot{\mathbf{a}}$ *will always refer to body (1)*, while $\mathbf{b}$, *etc.*, will refer to either of the other bodies (the massive ones). Also, from now on, we shall assume that body (1) has a *rigid motion* at the Newtonian approximation, so that its zero-order velocity field is

(2.5) $\qquad u^i = \dot{a}^i + \Omega_{ji}(x^j - a^j), \qquad \mathbf{x} \in D_1 \quad (\Omega_{ji} + \Omega_{ij} = 0).$

This assumption seems especially reasonable for a small body such as a spacecraft in free motion or a satellite, but it also seems hardly compatible with that saying that the body is made of a perfect fluid, since a small body cannot maintain its structure by its own gravitational field. However, the assumption of a perfect-fluid body, which has the advantage of making the calculations simpler, is unlikely to play an important role physically, because, as we shall see, the pressure plays no direct role in the final equations of motion.

---

[1]This implication is obvious for the integrals [7, Eqs. (4.6), (4.8) and (4.18)], for they depend linearly on the external potentials $U^{(a)}$, $B^{(a)}$ and $W^{(a)}$. Concerning the integral (4.10) of Ref. 7, which is reexpressed as Eq. (2.21) in Ref. 8, the implication follows from the fact that $k_{ij} = U h_{ij} = (U^{(a)} + u_a) h_{ij}$, with $h_{ij} = U_{,i} U_{,j}/(U_{,k} U_{,k})$ [7, Eq.(2.31)]. In body $(a)$ with $a > 1$, $h_{ij}$ is $O(\xi^0)$ while the contribution of body (1) to $U^{(a)}$ is $O(\xi^3)$.

# 3. EXPANSION OF THE INTEGRALS $I^1$, $J^1$, $K^1$ AS $\xi \to 0$

## 3.1 AUXILIARY EXPANSIONS

We shall give the 1PN correction to the acceleration of body (1) up to unknown $O(\xi)$ terms. In view of the equation for that correction, and since the Newtonian mass $M_1$ of (1) is ord($\xi^3$), this needs that we compute the foregoing integrals up to $O(\xi^4)$. From the assumption that $\rho$ is ord($\xi^0$) inside (1), it also follows for the self-part $u_1$ of the Newtonian potential that

(3.1) $$u_{1,i} = \text{ord}(\xi), \quad u_1 = \text{ord}(\xi^2)$$

inside body (1). Moreover, the spatial component of the zero-order expansion of the dynamical equation is just the Newtonian equation for a perfect fluid [6], and with (2.5) it gives [1, Eqs. (73.04,06,07)]:

(3.2) $$\rho\left[\ddot{a}^i + \left(\dot{\Omega}_{ji} - \Omega_{ik}\Omega_{jk}\right)\left(x^j - a^j\right)\right] - \rho\frac{\partial U}{\partial x^i} = -\frac{\partial p}{\partial x^i}, \quad \mathbf{x} \in D_1.$$

Inserting (2.3) and (2.4) into (3.2), we get:

(3.3) $$\rho\left(\dot{\Omega}_{ji} - \Omega_{ik}\Omega_{jk}\right)\left(x^j - a^j\right) - \rho u_{1,i} - \rho U^{(1)}_{,i,j}(\mathbf{a})\left(x^j - a^j\right) = -p_{,i} + O(\xi^2), \quad \mathbf{x} \in D_1,$$

and, accounting for $(3.1)_1$, we obtain inside (1): [2]

(3.4) $$p_{,i} = O(\xi).$$

Integrating this radially from the external surface of the body, we deduce

(3.5) $$p = O(\xi^2).$$

The expansion of the space tensor $h_{ij}$ [Note 1] is got from (2.3) and $(3.1)_1$:

(3.6) $$h_{ij}(\mathbf{x}) = \frac{U^{(1)}_{,i}(\mathbf{a})U^{(1)}_{,j}(\mathbf{a})}{U^{(1)}_{,k}(\mathbf{a})U^{(1)}_{,k}(\mathbf{a})} + O(\xi), \quad \mathbf{x} \in D_1.$$

This expansion with respect to $\xi$ contrasts with that with respect to the separation parameter $\eta$ between bodies, obtained in Ref. 10 [Eq. (5.21) there]. The simple but consequential point is that, inside (1) and as $\xi \to 0$, the principal part of the potential $U$ is the external potential $U^{(1)}$, whereas, inside (1) and as $\eta \to 0$, its principal part is the self-potential $u_1$. For this reason, the expansions as $\xi \to 0$ and those as $\eta \to 0$ are, generally speaking, incompatible. In particular, it does not make sense to take the point-particle limit ($\xi \to 0$) in the equations of motion obtained for well-separated bodies (*i.e.* in the equations of motion of the mass centers, as expanded for $\eta \to 0$).

---

[2] Together with the assumption that $\rho$ is ord($\xi^0$), this result constrains the state equation $\rho = F_\xi(p)$ in body (1), in particular $F_\xi$ must indeed depend on $\xi$.

The expansion of $k_{ij} = Uh_{ij}$ is obvious from $(3.1)_2$ and (3.6):

$$(3.7) \qquad k_{ij}(\mathbf{x}) = U^{(1)}(\mathbf{a}) \frac{U^{(1)}_{,i}(\mathbf{a})U^{(1)}_{,j}(\mathbf{a})}{U^{(1)}_{,k}(\mathbf{a})U^{(1)}_{,k}(\mathbf{a})} + O(\xi), \qquad \mathbf{x} \in D_1.$$

The expansion of $k_{jk,i}$, which contributes to the integral $K_4^{ai}$ [8, Eq. (2.21)], does not obtain by differentiating (3.7), because the small parameter $\xi$ determines the spatial scale inside (1), hence it does interfere with space differentiation. From $(3.1)_1$, one expects that

$$(3.8) \qquad u_{1,i,j} = \mathrm{ord}(\xi^0),$$

and this may be verified by writing the defining integral of $u_1$. E.g. for a homogeneous sphere,

$$(3.9) \qquad u_{1,i,j} = -4\pi G \rho \delta_{ij}/3.$$

Performing first the space differentiation of $Uh_{ij}$ and then inserting (2.3), (3.1) and (3.6,8) yields

$$(3.10) \quad k_{jk,i} - k_{ik,j} = U^{(1)}(\mathbf{a}) \left\{ \frac{g_j[u_{1,k,i} + U^{(1)}_{,k,i}(\mathbf{a})] - g_i[u_{1,k,j} + U^{(1)}_{,k,j}(\mathbf{a})]}{\mathbf{g}^2} \right.$$
$$\left. + \frac{2g_lg_k[g_i(u_{1,l,j} + U^{(1)}_{,l,j}(\mathbf{a})) - g_j(u_{1,l,i} + U^{(1)}_{,l,i}(\mathbf{a}))]}{(\mathbf{g}^2)^2} \right\} + O(\xi),$$

in which $\mathbf{g} \equiv \nabla U^{(1)}(\mathbf{a})$ is the external Newtonian gravity acceleration.

## 3.2 INTEGRAL $\mathbf{I}^1$

We have from (2.5) and the definition of $\mathbf{I}^1 = (I^{1i})$ [7, Eq. (4.8)]:

$$(3.11) \qquad I^{1i} = \dot{a}^i \int_{D_1} [p + \rho(\mathbf{u}^2/2 + \Pi + U)] \, dV + O(\xi^4).$$

Using again the definition of $I^{1i}$, plus the definition of the Newtonian mass center $\mathbf{a}$, and also $(3.1)_1$, (3.5) and (A13), we have

(3.12) $\int_{D_1} [p + \rho(\mathbf{u}^2/2 + \Pi + U)] dV = M_a \dot{\mathbf{a}}^2/2 + \int_{D_1} \rho U^{(1)} dV + O(\xi^5).$

From (2.3), we get by using (3.11) and (3.12):

(3.13) $\mathbf{I}^1 = M_a [\dot{\mathbf{a}}^2/2 + U^{(1)}(\mathbf{a})]\dot{\mathbf{a}} + O(\xi^4).$

## 3.3 INTEGRAL $\mathbf{J}^1$

The Taylor expansion (2.3) gives

(3.14) $\int_{D_1} \sigma_1 U^{(1)}{}_{,i} dV = \alpha_1 U^{(1)}{}_{,i}(\mathbf{a}) + \beta_{1j} U^{(1)}{}_{,i,j}(\mathbf{a}) + O(\xi^5),$

where

(3.15) $\alpha_b \equiv \int_{D_b} \sigma_1 dV, \quad \beta_{bi} \equiv \int_{D_b} \sigma_1(\mathbf{x})(x^i - b^i) dV(\mathbf{x}) \quad (b = 1,\ldots, N),$

and, accounting for $(2.2)_1$ in the definition of $\sigma_1$ [7, Eq. (2.24)], we get as for (3.12):

(3.16) $\alpha_1 = M_1[\dot{\mathbf{a}}^2/2 + U^{(1)}(\mathbf{a})] + M_1^1 + O(\xi^5),$

(3.17) $\beta_{1j} = M_1^1 (a_1^j - a^j) + O(\xi^5),$

(3.18) $M_1^1 = M_1[\dot{\mathbf{a}}^2/2 + U^{(1)}(\mathbf{a})]_{T=0} + O(\xi^5)$

(the latter is the 1PN correction to the mass of body (1) [7, Eq. (3.7)].

In the same way, we have

(3.19) $\int_{D_1} \rho B^{(1)}{}_{,i} dV = M_1 B^{(1)}{}_{,i}(\mathbf{a}) + O(\xi^5),$

and $B^{(1)}{}_{,i}(\mathbf{a})$ [which is ord($\xi^0$)] can be obtained by differentiating term by term the series

(3.20) $\dfrac{B^{(1)}(\mathbf{X})}{G} = \sum_{b=2}^{N} \int_{D_b} \dfrac{\sigma_1(\mathbf{x})}{|\mathbf{X} - \mathbf{x}|} dV(\mathbf{x}) = \sum_{b=2}^{N} \left\{ \dfrac{\alpha_b}{|\mathbf{X} - \mathbf{b}|} + \dfrac{X^j - b^j}{|\mathbf{X} - \mathbf{b}|^3} \beta_{bj} + \dfrac{(X^j - b^j)(X^k - b^k)}{2|\mathbf{X} - \mathbf{b}|^5} D_{bjk} + \ldots \right\}$

with

$$\text{(3.21)} \quad D_{bjk} \equiv \int_{D_b} \sigma_1(\mathbf{x})\left[3(x^j - b^j)(x^k - b^k) - |\mathbf{x} - \mathbf{b}|^2 \delta_{jk}\right] dV(\mathbf{x}).$$

(In (3.20), the series inside the braces is the standard multipole expansion of a Newtonian potential, so we admit that differentiation and series summation do commute here. Then, we do not need a small parameter $\eta$ of ``good separation''. In practice, already the quadrupole term is likely to be extremely small.) The last term in the integral $J^{ai}$ [7, Eq. (4.18)] is [8, appendices]:

$$\text{(3.22)} \quad \int_{D_1} \rho(\partial^3 W/\partial x^i \partial T^2) dV = \int_{D_1} \rho(\partial^3 W^{(1)}/\partial x^i \partial T^2) dV + (d/dT) \int_{D_1} \rho(\partial^2 w_1/\partial x^i \partial T) dV + O(1/c^2).$$

The definition of $W$ [7, Eq. (4.14)] implies that here

$$\text{(3.23)} \quad \frac{\partial^2 w_1}{\partial x^i \partial T} = \frac{G}{2} \frac{\partial}{\partial T} \int_{D_1} |\mathbf{x} - \mathbf{y}| \partial_i \rho(\mathbf{y}, T) dV(\mathbf{y})$$

$$= \frac{G}{2} \left( \int_{\partial D_1} |\mathbf{x} - \mathbf{y}| \partial_i \rho(\mathbf{y}, T) \mathbf{u}.\mathbf{n} dS(\mathbf{y}) - \int_{D_1} |\mathbf{x} - \mathbf{y}| u^j(\mathbf{y}, T) \frac{\partial^2 \rho}{\partial y^i \partial y^j} dV(\mathbf{y}) \right),$$

since the assumed rigid motion implies that $\psi \equiv \partial_i \rho$ verifies $\partial_T \psi = -u^j \partial_j \psi$. From this, and assuming that $\partial_i \rho = O(\xi^{-1})$ and $\partial_i \partial_j \rho = O(\xi^{-2})$ [this is true even in the severe case where the dependence of $\rho$ with $\xi$ is homothetical, $\rho^{(\xi)}(\mathbf{x}) = \rho^{(1)}[\mathbf{a}^{(1)} + (\mathbf{x} - \mathbf{a}^{(\xi)})/\xi]$ for $\mathbf{x} \in D_1$], it follows immediately that

$$\text{(3.24)} \quad \int_{D_1} \rho(\partial^2 w_1/\partial x^i \partial T) dV = O(\xi^5).$$

On the other hand, we get by the usual Taylor expansion:

$$\text{(3.25)} \quad \int_{D_1} \rho(\partial^3 W^{(1)}/\partial x^i \partial T^2) dV = M_1 (\partial^3 W^{(1)}/\partial x^i \partial T^2)(\mathbf{a}) + O(\xi^5).$$

The value of the external field on the r.h.s. may be obtained by writing $W^{(1)}$ as a multipole series [1, Eq. (76.25)], like (3.20). The value of (3.25) corresponding to the two first terms in this series was given previously [8b, Eq. (A24)]. In the observationally relevant case that the zero-order density $\rho$ is spherical in the ``large'' bodies, their inertia tensors are time-independent and we get thus:

$$\text{(3.26)} \quad \int_{D_1} \rho(\partial^3 W^{(1)}/\partial x^i \partial T^2) dV = -\frac{GM_1}{2} \sum_{b=2}^{N} M_b \left( \ddot{b}^k \frac{\partial^2 |\mathbf{a} - \mathbf{b}|}{\partial a^i \partial a^k} - \dot{b}^k \dot{b}^j \frac{\partial^3 |\mathbf{a} - \mathbf{b}|}{\partial a^i \partial a^k \partial a^j} \right) + ... + O(\xi^5),$$

where the points of suspension indicate quantities coming from terms of rank > 2 in the multipole series expansion of $W^{(1)}$. Inserting (3.14), (3.19) and (3.26) into [7, Eq.(4.18)], we get:

$$(3.27) \quad J^{1i} = \alpha_1 U^{(1)}{}_{,i}(\mathbf{a}) + \beta_{1j} U^{(1)}{}_{,i,j}(\mathbf{a}) - \frac{GM_1}{2} \sum_{b=2}^{N} M_b \left( \ddot{b}^k \frac{\partial^2 |\mathbf{a}-\mathbf{b}|}{\partial a^i \partial a^k} - \dot{b}^k \dot{b}^j \frac{\partial^3 |\mathbf{a}-\mathbf{b}|}{\partial a^i \partial a^k \partial a^j} \right) + \ldots$$

$$+ GM_1 \left( \frac{\partial}{\partial X^i} \right)_{\mathbf{X}=\mathbf{a}} \sum_{b=2}^{N} \left\{ \frac{\alpha_b}{|\mathbf{X}-\mathbf{b}|} + \frac{X^j - b^j}{|\mathbf{X}-\mathbf{b}|^3} \beta_{bj} + \frac{(X^j - b^j)(X^k - b^k)}{2|\mathbf{X}-\mathbf{b}|^5} D_{bjk} + \ldots \right\} + O(\xi^5).$$

## 3.4 INTEGRAL $\mathbf{K}^1$

Using (2.3), (3.4) and (3.7) into [8, Eq.(2.21)] gives immediately

$$(3.28) \quad K_1^{1i} = M_1 \, U^{(1)}(\mathbf{a}) \, U^{(1)}{}_{,i}(\mathbf{a}) + O(\xi^4).$$

Also, (3.5) implies that

$$(3.29) \quad K_2^{1i} = O(\xi^5),$$

whereas Eqs. (2.3) and (3.1) give us

$$(3.30) \quad K_3^{1i} = -2 M_1 \, U^{(1)}(\mathbf{a}) \, U^{(1)}{}_{,i}(\mathbf{a}) + O(\xi^4),$$

and (2.5) plus (3.7) lead to

$$(3.31) \quad K'^{1i} = M_1 \, \dot{\mathbf{a}} \cdot \nabla U^{(1)}(\mathbf{a}) \, U^{(1)}(\mathbf{a}) \, U^{(1)}{}_{,i}(\mathbf{a}) / (\nabla U^{(1)}(\mathbf{a}))^2 + O(\xi^4).$$

Finally, we get from (2.5) and (3.10):

$$(3.32) \quad K_4^{1i} = M_1 U^{(1)}(\mathbf{a}) \dot{a}^j \dot{a}^k \left\{ \frac{g_j \left[ S_{ki} + U^{(1)}{}_{,k,i}(\mathbf{a}) \right] - g_i \left[ S_{kj} + U^{(1)}{}_{,k,j}(\mathbf{a}) \right]}{\mathbf{g}^2} \right.$$
$$\left. + \frac{2 g_l g_k \left[ g_i \left( S_{lj} + U^{(1)}{}_{,l,j}(\mathbf{a}) \right) - g_j \left( S_{li} + U^{(1)}{}_{,l,i}(\mathbf{a}) \right) \right]}{(\mathbf{g}^2)^2} \right\} + O(\xi^4),$$

where the (symmetric) structure-dependent space tensor $\mathbf{S}$ is given by

$$(3.33) \quad S_{kj} \equiv \int_{D_1} \rho \, u_{1,k,j} \, dV / M_1 \equiv \int_{D_1} \rho \, u_{1,k,j} \, dV / \int_{D_1} \rho \, dV.$$

From the examination of the 1PN correction to the equation of motion [7, Eq. (4.9)] and the explicit values of the corresponding integrals [Eqs. (3.13), (3.27), (3.28-32)], one sees that the only terms which make the structure of the small body play a role in the motion of its own mass center are precisely those terms of (3.32) that include the $\mathbf{S}$ tensor.

# 4. THE CASE WITH ONE MASSIVE BODY, ASSUMED STATIC AND SPHERICAL

Assume that (2) is the only ``massive'' body, that it is at rest at the origin (in the preferred frame), and that the Newtonian density $\rho$ inside (2) is spherically symmetrical. Set

(4.1) $\quad\quad m \equiv M_1, \quad M \equiv M_2, \quad \mathbf{v} = \dot{\mathbf{a}}, \quad r \equiv |\mathbf{a}|, \quad \mathbf{n} \equiv \mathbf{a}/r, \quad \mathbf{x}_{11} \equiv c^2(\mathbf{a}_{(1)} - \mathbf{a}).$

We may omit the remainders, once we remember that the 1PN correction to the Newtonian acceleration of the small body shall be obtained up to unknown $O(\xi)$ terms. We get from (3.13):

(4.2) $$\dot{\mathbf{I}}^1 = -m\frac{GM}{r^2}\left[\left(\frac{\mathbf{v}^2}{2} + \frac{GM}{r}\right)\mathbf{n} + 2(\mathbf{v}\cdot\mathbf{n})\mathbf{v}\right].$$

In the expression (3.27) for $J^{1i}$, we have from [7, Eq. (3.15)], and from Eqs. (3.16) and (3.17) here, in the notation (4.1):

(4.3) $\quad\quad \alpha_1 = m\{[\mathbf{v}^2/2 + GM/r]_T + [\mathbf{v}^2/2 + GM/r]_{T=0}\}, \quad \beta_{1j} = M_1^{\ 1}(a_1^j - a^j) = m\,x_{11}^{\ j}.$

Let us compute $\alpha_2$. We have from the definition of $\sigma_1$ [7, Eq. (2.24)]:

(4.4) $$\alpha_2 = M_2^{\ 1} + \int_{D_2} \rho\,(\mathbf{u}^2/2 + \Pi + U)\,dV.$$

By $(2.2)_1$ and [7, Eq. (2.23)], and since $U^{(2)} = 0$ [or more precisely $U^{(2)} = O(\xi^3)$], we get:

(4.5) $$M_2^{\ 1} = 2\varepsilon_2, \quad \varepsilon_a \equiv \int_{D_a} \rho\,u_a\,dV/2.$$

From (A3) [which applies here, not with an $O(\eta^3)$ remainder, but with an $O(\xi^3)$ one, because the body (2) is alone apart from the small body (1)], we get as for (A12)

(4.6) $$\rho\Pi + p = \rho V_2 = \rho u_2 \quad\quad (\Omega_2 = 0)$$

inside (2). Hence, using Fock's [1] integral (74.24):[3]

---

[3] We checked the full derivation of this integral. Apart from exact mathematical manipulations, it uses ``Liapunov's equation'' (A4) and is hence valid up to an $O(\eta^3)$ remainder for well-separated bodies, and up to an $O(\xi^3)$ remainder (for $b = 2$) in the present case.

(4.7) $$3 \int_{D_a} p \, dV = \varepsilon_a - 2 T_a, \qquad T_a \equiv \int_{D_a} \rho \Omega_a dV,$$

we get here by (4.6)

(4.8) $$\int_{D_2} \rho (\mathbf{u}^2/2 + \Pi + u_2) \, dV = \int_{D_2} (2\rho u_2 - p) \, dV = (11/3) \, \varepsilon_2.$$

Since $U^{(2)} = 0$, we get from (4.4-5) and (4.8):

(4.9) $$\alpha_2 = (17/3)\varepsilon, \qquad \varepsilon \equiv \varepsilon_2 \equiv \int_{D_2} \rho u_2 \, dV/2.$$

The higher-order terms in (3.20): $\beta_{2j}$, $D_{2jk}$, etc., are zero: using again the fact that $U^{(2)} = 0$, Eq. (2.2) and [7, Eqs. (2.23-24)] imply that $\sigma_1$, which is time-independent here, is

(4.10) $$\sigma_1 = \rho(\Pi + 2u_2).$$

Just like $\rho$, and hence also $\Pi$ and $u_2$, this is spherically symmetrical around **b**. Hence indeed the definitions (3.15)$_2$ and (3.21) give immediately

(4.11) $$\beta_{2j} = 0, \qquad D_{2jk} = 0.$$

The same is true for the terms omitted in the series (3.20), because the external Newtonian potential of a spherical distribution is exactly given by the monopole term. Moreover, the integral on the r.h.s. of (3.25) is zero in the present static case. Accounting for these results and for (4.3) and (4.9), and using the Newtonian energy integral: $GM/r - \mathbf{v}^2/2 = \text{Const.}$, Eq. (3.27) gives us

(4.12) $$\mathbf{J}^1 = m \frac{GM}{r^2} \left[ -\left(\mathbf{v}^2 + 2\frac{GM}{r_{T=0}} + \frac{17}{3}\frac{\varepsilon}{M}\right)\mathbf{n} + \frac{3}{r}(\mathbf{x}_{11}.\mathbf{n})\mathbf{n} - \frac{\mathbf{x}_{11}}{r} \right].$$

As to the integral $\mathbf{K}^1$, we get easily from Eqs. (3.28-32):

(4.13) $$\mathbf{K}_1^1 = -m\frac{GM}{r^2}\left(\frac{GM}{r}\mathbf{n}\right), \qquad \mathbf{K}_2^1 = \mathbf{0}, \qquad \mathbf{K}_3^1 = 2m\frac{GM}{r^2}\left(\frac{GM}{r}\mathbf{n}\right),$$

(4.14) $$\dot{\mathbf{K}}^{'1} = m\frac{GM}{r^2}\left[\left(-3(\mathbf{v}.\mathbf{n})^2 + \mathbf{v}^2 - \frac{GM}{r}\right)\mathbf{n} + (\mathbf{v}.\mathbf{n})\mathbf{v}\right],$$

(4.15) $$\mathbf{K}_4^1 = \mathbf{K}_S + m\frac{GM}{r^2}\left[\left(-\mathbf{v}^2\right)\mathbf{n} + (\mathbf{v}.\mathbf{n})\mathbf{v}\right],$$

(4.16)  $\mathbf{K_S} \equiv -mr\{(\mathbf{v.n})\mathbf{Sv} - (\mathbf{Sv.v})\mathbf{n} + 2(\mathbf{v.n})[(\mathbf{Sv.n})\mathbf{n} - (\mathbf{v.n})\mathbf{Sn}]\}$  $(\mathbf{Sv} \equiv (S_{ij} v^j))$.

Inserting (3.18), (4.2), (4.12), (4.13-15) into the equation for 1PN correction [7, Eq. (4.9)], and using again the integral $GM/r - \mathbf{v}^2/2 = \text{Const.}$, gives the coefficient of $1/c^2$ in the acceleration of the small body:

$$(4.17) \quad \frac{d\mathbf{v}_{11}}{dT} = \frac{\mathbf{K_S}}{m} + \frac{GM}{r^2}\left\{\left(\frac{2GM}{r} - \frac{17\varepsilon}{3M} + 3\left[(\mathbf{v.n})^2 + \frac{\mathbf{x}_{11}.\mathbf{n}}{r}\right] - 2\mathbf{v}^2\right)\mathbf{n} + 2(\mathbf{v.n})\mathbf{v} - \frac{\mathbf{x}_{11}}{r}\right\} + O\left(\frac{1}{c^2}\right) + O(\xi).$$

This is *exactly* the coefficient of $1/c^2$ in the acceleration of a test particle in a spherical static field according to the scalar theory [9, Eqs. (22) and (36)],[4] *except for the structure-dependent term* $\mathbf{K_S}/m$ [and up to the $O(\xi)$ remainder]. Thus, in the case of just one massive body, assumed spherical and static, the 1PN acceleration of a small body in the point-particle limit differs from the 1PN acceleration of a test particle just by the complementary acceleration

(4.18)  $\mathbf{A_S} = -r\{(\mathbf{v.n})\mathbf{Sv} - (\mathbf{Sv.v})\mathbf{n} + 2(\mathbf{v.n})[(\mathbf{Sv.n})\mathbf{n} - (\mathbf{v.n})\mathbf{Sn}]\}/c^2$,

that should be felt, according to the present scalar theory, by a real small body.

## 5. THE VIOLATION OF THE WEAK EQUIVALENCE PRINCIPLE AND THE REASON FOR IT

We come back to the case of a *general 1PN system of self-gravitating bodies* ($b$) ($b = 2,..., N$) and note that the structure-dependent part of the acceleration of the small body (1) is [Eq. (3.32)]:

$$(5.1) \quad \mathbf{A_S} = \frac{\left(\mathbf{K}_4^1\right)_\mathbf{S}}{c^2 M_1} = \frac{U^{(1)}(\mathbf{a})}{c^2 \mathbf{g}^2}\{(\mathbf{v.g})\mathbf{Sv} - (\mathbf{Sv.v})\mathbf{g} + 2\frac{\mathbf{v.g}}{\mathbf{g}^2}[(\mathbf{Sv.g})\mathbf{g} - (\mathbf{v.g})\mathbf{Sg}]\}.$$

By analyzing the difference between the general case and the particular case investigated in Section 5, it appears almost certain that the acceleration in the point-particle limit of the general case also differs from the acceleration of a test particle *only* by the structure-dependent acceleration, thus only by $\mathbf{A_S}$. This could be checked by appealing to the 1PN equation of motion for a test particle in the scalar theory, which has been given in Ref. 15, p. 22.

---

[4] In general relativity, the equation of a test particle in a spherical static field is the same, apart from the coefficient 5/3 instead of 17/3, due to the fact that the PN correction to the active mass is $(5/3)\varepsilon/c^2$ instead of $(17/3)\varepsilon/c^2$ [9, Eq. (19)]. Note that it is the *total* active mass which determines the orbits and which is thus ``measured", hence that difference in the coefficients is not testable.

In the case that the small body is spherical, we may evaluate easily the **S** tensor (3.33):

(5.2)    $S_{ij} = -4\pi G \tilde{\rho}\, \delta_{ij}/3$,    $\tilde{\rho} \equiv \int_0^{r_1} 4\pi r^2 \rho^2\, dr/m = \int_{D_1} \rho^2\, dV/\int_{D_1} \rho\, dV$.

(Thus, $\tilde{\rho} = \rho$ for a homogeneous sphere.) The complementary acceleration (5.1) is then:

(5.3)    $$\mathbf{A_S} = \frac{4\pi G \tilde{\rho}}{3} \frac{U^{(1)}(\mathbf{a})}{c^2 \mathbf{g}^2}[\mathbf{v}^2 \mathbf{g} - (\mathbf{v}.\mathbf{g})\, \mathbf{v}].$$

Whether the small body (1) is spherical or not, if it goes to a large distance $R$ from the system of massive bodies, then $\mathbf{A_S}$ increases like $R$ ! However, for the expansion at small $\xi$, which has been used, the principal part of the Newtonian potential is the external potential $U^{(1)}$, which is *a priori* independent of $\xi$ [see Eqs. (2.3) and (3.1), and see after Eq. (3.6)]. But if the small body goes too far, the self-potential $u_1$ does no longer stay negligible with respect to $U^{(1)}$, so that the expansion does not make sense any more. Anyhow, as all post-Newtonian expansions, the present expansions are valid *in the near zone,* since the distances are assumed $\mathrm{ord}(\lambda^0)$ at the stage of writing the general PN expansions, where $\lambda$ is the field-strength parameter. Moroever, the distances have been implicitly assumed $\mathrm{ord}(\xi^0)$ in the point-particle limit. The latter point is that which makes $U^{(1)}$ independent of $\xi$. Thus, the linear increase with $R$ is predicted for *bounded* distances $R$, such that $u_1$ is still negligible with respect to $U^{(1)}$. Note that $\nabla u_1$ must also remain negligible with respect to $\nabla U^{(1)}$, and this is the most stringent condition.

Now *how can a such irreducible theoretical difference exist between the acceleration of the mass center of a small extended body and the acceleration of a test particle?* The general reason that makes it *a priori* possible is simply that the small body is gravitationally active and the theory is nonlinear! Due to our definition [7] of the mass center by averaging the rest-mass density, which is conserved at the first PNA *i.e.* up to $O(c^{-4})$ terms [6], the 1PN acceleration of the mass center is just the barycenter of the acceleration field inside the body:

(5.4)    $$M_a^{(1)} \ddot{\mathbf{a}}_{(1)} = \int_{D_a} \rho_{(1)} \dot{\mathbf{u}}_{(1)}\, dV + O(c^{-4}).$$

The 1PN acceleration field $\dot{\mathbf{u}}_{(1)}$, more precisely the 1PN correction $\dot{\mathbf{u}}_1/c^2$, depends nonlinearly on the matter fields (directly and *via* the gravitational potentials $U$ and $A$). However, the zero-order part $\dot{\mathbf{u}}$ does depend linearly on the zero-order density $\rho$ (and only on $\rho$). It is the linearity of the dependence of $\dot{\mathbf{u}}$ on $\rho$ that allows the very *definition* of the actio-reactio principle for Newtonian gravity (NG) – and it turns out that NG does obey this principle, essentially because in NG the force depends only on the current positions. This principle, involving this linearity as a necessary condition, is the deep reason why the integral of the Newtonian self-force is zero. This finally results in Eq. (2.4), which shows that a small body does behave like a test particle as regards the Newtonian part of its acceleration. As to the field $\dot{\mathbf{u}}_1$, however, since it depends nonlinearly on (all) matter fields, we cannot even *define* which part of $\dot{\mathbf{u}}_1(\mathbf{x})$ comes from the body itself where **x** belongs, and which part comes from the other bodies. *Via* the linear average (5.4), this nonlinear dependence is carried over to the 1PN correction to the acceleration of the mass center. Thus, there is no *a priori* reason why the acceleration of the mass center should depend only on the other bodies, even if the body is as small as desired. Now it is the $\mathbf{K}^a$ integral [8, Eq. (2.21)] that keeps the main nonlinearities. This integral depends on the 1PN part of the spatial metric (the space tensor **k**), which is influenced by the body (1) itself whose motion is sought for. Even more particularly, the $\mathbf{K}_4^a$ integral depends on the derivatives $k_{jk,i}$ of the spatial metric: *the specific reason why the influence of the small body (1) does indeed survive in the point-particle limit is that the 1PN spatial metric contains first-order spatial derivatives of the Newtonian potential U*, so that the Christoffel symbols (or the derivatives $k_{jk,i}$) contain *second* derivatives of *U*, whose self-part is independent of the size of the body, Eq. (3.8).

Yet the present form of the 1PN spatial metric, $g_{ij} = \delta_{ij} + 2k_{ij}/c^2 + O(c^{-4})$, where $k_{ij} = Uh_{ij}$, which thus involves derivatives $U_{,i}$ [see Note 1], *is not specific to the investigated scalar theory, because it occurs also for the standard Schwarzschild metric:* the spatial part of the latter may also be written in just the same form, specifically $U = GM/r$ in that case [9, Eq. (26bis)]. We note that the usual gauge conditions used in the PNA of GR are close to the harmonic gauge, and give 1PN spatial metrics of the form $g_{ij} = \delta_{ij}(1+2U/c^2)$, thus without any derivative $U_{,i}$ [1-2, 4]. But since Schwarzschild's metric is an exact solution of GR (which is indeed often used as an astronomically relevant simplification for PN calculations), we may confidently conjecture that, for some gauge conditions, the 1PN metric of GR should have a similar form and contain derivatives $U_{,i}$. Hence, in these gauges (and following the asymptotic PNA), the acceleration in the point-particle limit should also differ from the acceleration of a test particle and contain a structure-dependent term – which would be a violation of the weak equivalence principle (WEP).

However, some clarification seems necessary, regarding what may be called a violation of the WEP in a relativistic theory of gravitation.[5] The violation revealed in the scalar theory by Eqs. (4.18) or (5.1), appears only for the (1PN) ``coordinate acceleration'' $\dot{\mathbf{u}}_{(1)}$,[6] and only for an extended body: the local dynamical equation of the scalar theory, which is at the basis of the present work, is

(5.5) $\quad T_\mu{}^\nu{}_{;\nu} = b_\mu, \qquad b_0(\mathbf{T}) \equiv \frac{1}{2} g_{jk,0} T^{jk}, \qquad b_i(\mathbf{T}) \equiv -\frac{1}{2} g_{ik,0} T^{0k}$

(**T** is the energy-momentum tensor). Although this is a preferred-frame equation, it makes gravitation just as ``universal'' as it is with $T_\mu{}^\nu{}_{;\nu} = 0$, which is the corresponding equation of GR. In fact, it is easy to check that the difference between the two equations has no influence on the presence or absence of the structure-dependent acceleration (5.1). Again, what makes the latter arise is just the fact that the 1PN spatial metric depends on the spatial derivatives of the Newtonian potential – a fact that also should apply to GR in some gauges. It is also worth to mention that the exact equation of motion for a test particle in the scalar theory, which is formulated in terms of the ``effective'' (curved) metric, involves the *exact* identity between inertial mass and (passive) gravitational mass: both are $m(v) \equiv m(0).(1 - v^2/c^2)^{-1/2}$, the velocity being measured with a local time $t_x$ and its modulus $v$ being defined with the spatial metric **g** in the preferred frame E [see Ref. 5, Eq. (12)]. The gravitational force is $m(v)\mathbf{g}_{exact}$ with $\mathbf{g}_{exact}$ the gravity acceleration, that depends only on the position in the frame E.

---

[5] We mean a theory with a correct Newtonian limit, accounting for special relativity, and reducing to it if $G = 0$ : the present theory is one such theory, although it predicts preferred-frame effects.

[6] In the case of the present theory, this acceleration is calculated with the preferred time and using Cartesian coordinates bound to the preferred frame, so that it has an absolute meaning [5].

One may add some remarks that might be relevant to the analysis of the tests of the WEP. When one calculates, already for a test particle, the coordinate acceleration **A**, thus now in terms of Cartesian space coordinates (bound to E) and in terms of the preferred time *T*, one finds that **A** depends on the position *and* on the coordinate velocity [5, Eq. (18)]. (The acceleration in a reference frame bound to the mass center of the solar system may then be calculated by Lorentz transform [5,15].) *Just the same occurs in GR, i.e.* there also the coordinate acceleration of a test particle depends on the position and on the coordinate velocity [16, Eq. (9.1.2), and see Eq. (4.17) above]. The mere difference is that, in GR, there is *a priori* much more latitude for choosing the coordinate system. However, if one aims at testing the WEP at relative precisions of $10^{-12}$ or better, one should try considering a coordinate system adapted to precise PN calculations, in order that one may check the effect of both the Newtonian acceleration due to the Sun and planets, and the PN part of the acceleration, none of which is *a priori* negligible at this level. But such coordinates adapted to precise PN calculations in the solar system are simply inertial coordinates, thus are coordinates bound to the mass center of the solar system, and whose axes do not rotate with respect to distant astronomical pointers [17-19]. Thus, for that matter, the difference between GR and the scalar theory is just that, in the latter, one has to account for the ``absolute'' velocity of the mass center of the solar system. However, from a preliminary adjustment (neglecting the self-rotation of the planets) [12, 20] of the equations of motion for the mass centers of the scalar theory to the DE403 ephemeris [21], we find that the (best-fitting) velocity **V** of the mass center of the solar system through the preferred frame of the scalar theory is likely to be of the order of a few km/s [12, 20]. This is not negligible, but on the other hand it means that the velocities of a celestial body like the Earth, relative to either the preferred frame of the scalar theory or the inertial frame that is bound to the mass center of the solar system, both should have more or less the same order of magnitude. Note that, in a laboratory on Earth, the velocity relative to the latter inertial frame has a daily variation, hence cannot be considered constant. *In summary,* relativistic theories of gravitation predict velocity-dependent accelerations for which the relevant velocity has astronomical magnitude and has a daily variation in a laboratory frame. Hence it would seem worth to perform a detailed analysis of tests of the WEP in the framework of PN calculations in the solar system, in order to check whether one really may content oneself with a Newtonian analysis at least for the test-particle acceleration (as is systematically done in the discussion of the torsion-balance tests, see Ref. 22 and references therein). Moreover, in GR also (as in the scalar theory, but, in GR, depending on the gauge condition used), one may expect that a structure-dependent acceleration could be present if the finite extension of the body were taken into account within the asymptotic PNA scheme.

# 6. ORDER-OF-MAGNITUDE ESTIMATES. COMPARISON WITH THE PIONEER ANOMALY

The structure-dependent part $\mathbf{A_S}$ of the acceleration of the small body has the form

(6.1) $\qquad \mathbf{A_S} = A_{S\,max}\,\mathbf{w}, \qquad A_{S\,max} \equiv \dfrac{4\pi G \tilde{\rho}}{3} \dfrac{\mathbf{v}^2}{c^2} \dfrac{U^{(1)}(\mathbf{a})}{|\mathbf{g}|}$

where $\tilde{\rho}$ is a kind of average density (in particular, $\tilde{\rho} = \rho$ if the small body is homogeneous):

(6.2) $\qquad \tilde{\rho} \equiv -(\mathrm{tr}\mathbf{S})/4\pi G = \int_{D_1} \rho^2\,dV / \int_{D_1} \rho\,dV,$

and $\mathbf{w}$ is a space vector depending on the relative orientation of the gravity acceleration $\mathbf{g}$ and the velocity $\mathbf{v}$, and such that $|\mathbf{w}|$ is at most of the order of unity: if the small body is spherical,

(6.3) $\qquad \mathbf{w} \equiv \mathbf{g'} - (\mathbf{v'}\cdot\mathbf{g'})\,\mathbf{v'}, \qquad \mathbf{g'} \equiv \mathbf{g}/|\mathbf{g}|, \qquad \mathbf{v'} \equiv \mathbf{v}/|\mathbf{v}|,$

so that indeed $|\mathbf{w}| \leq 1$ in that case. In the general case, $\mathbf{w}$ depends on the geometry of the body:

(6.4) $\qquad \mathbf{w} \equiv -(\mathbf{v'}\cdot\mathbf{g'})\mathbf{S'v'} + (\mathbf{S'v'}\cdot\mathbf{v'})\mathbf{g'} - 2(\mathbf{v'}\cdot\mathbf{g'})[(\mathbf{S'v'}\cdot\mathbf{g'})\mathbf{g'} - (\mathbf{v'}\cdot\mathbf{g'})\mathbf{S'g'}],$

where the adimensional symmetric operator $\mathbf{S'}$ verifies $\mathrm{tr}\mathbf{S'} = 3$ and is given by

(6.5) $\qquad S'_{ij} \equiv 3\int_{D_1} \rho\, u_{1,i,j}\,dV / \int_{D_1} \rho(-4\pi G\rho)\,dV$

(thus $\mathbf{S'} = \mathbf{1}$ in the spherical case).

Let us assess the *maximum* magnitude $A_{S\,max}$ for a small body on the surface of the Earth. The external potential $U^{(1)}(\mathbf{a})$ is dominated by the potential due to the Sun, which is $G = GM_\odot \approx 3\times 10^{-4}$ in the IAU system. On the other hand, the external gravity acceleration $\mathbf{g}$ comes essentially from the contribution of the Earth. With $GM_\oplus \approx 9\times 10^{-10}$ and $r_\oplus \approx 6.3/(1.5\times 10^5)$ (IAU), we get $U^{(1)}(\mathbf{a})/|\mathbf{g}| \approx 6\times 10^{-4}$ au $\approx 9\times 10^7$ m. We have $4\pi G\rho/3 \approx 4\times 7\times 10^{-11}\times 10^3 d$ (MKS), where $d$ is the density in g/cm$^3$. Finally, as mentioned in the last paragraph of Sect. 6, we find from ephemerides adjustment that the velocity $v$ (which, in the scalar theory, is relative to the preferred frame) should have the same order of magnitude as the orbital velocity of the Earth ($\approx$ 30 km/s), thus $v^2/c^2 \approx 10^{-8}$. Entering these numbers into (6.1) yields

(6.6) $\qquad A_{S\,max} \approx 2\times 10^{-7} d$ m/s$^2$ on the surface of the Earth.

Thus, $A_{S\,max} \approx 2 \times 10^{-8} g$ for $d \approx 1$. The modulus $|\mathbf{A_S}|$ may be much smaller than this, depending on the relative orientations of $\mathbf{v}$, $\mathbf{g}$, and the principal axes of the structure tensor $\mathbf{S'}$. Also note that, for two different spherical bodies, the two values of $\mathbf{A_S}$ would differ only due to the different densities. Hence, although the maximum order of magnitude (6.6) seems dangerous, a rather detailed analysis would be needed to see whether this is allowed, or not, by past experiments that aimed at testing the weak equivalence principle. (See also the discussion at the end of Sect. 6.)

As to the Pioneer spacecrafts [14], they are not spherical, but, for an order-of-magnitude estimate, we may assume that one principal axis of the $\mathbf{S}$ (or $\mathbf{S'}$) tensor is the spin axis, with unit vector $\mathbf{n}$, and that the eigenvalues in the axes perpendicular to $\mathbf{e}_3 \equiv \mathbf{n}$ are equal, thus $(S'_{ij}) =$ diag$(s_1, s_1, s_2)$. Decomposing $\mathbf{v'} = \mathbf{v'}_\perp + v'_{/\!/}\, \mathbf{n}$, $\mathbf{g'} = \mathbf{g'}_\perp + g'_{/\!/}\, \mathbf{n}$ (with $\mathbf{n}.\mathbf{v'}_\perp = \mathbf{n}.\mathbf{g'}_\perp = 0$), we get

(6.7) $\quad \mathbf{w} = (s_1\, \mathbf{v'}_\perp^2 + s_2\, v'^2_{/\!/})\mathbf{g'}_\perp - s_1(\mathbf{v'}.\mathbf{g'})\mathbf{v'}_\perp + [(s_1\, \mathbf{v'}_\perp^2 + s_2\, v'^2_{/\!/})\, g'_{/\!/} - s_2(\mathbf{v'}.\mathbf{g'})v'_{/\!/}]\mathbf{n}$

$\qquad + 2(\mathbf{v'}.\mathbf{g'})(s_1 - s_2)[\, v'_{/\!/}\, g'_{/\!/}\, \mathbf{g'}_\perp - (\mathbf{v'}_\perp.\mathbf{g'}_\perp)g'_{/\!/}\, \mathbf{n}].$

In particular, if both $\mathbf{g}$ and $\mathbf{v}$ are parallel to the spin axis $\mathbf{n}$ (which does not seem far from being true if one takes for $\mathbf{v}$ the heliocentric velocity [14]), then $\mathbf{w}$ and hence $\mathbf{A_S}$ are exactly zero. It seems that a more realistic case is when $\mathbf{g}$ is parallel to $\mathbf{n}$ and in the opposite direction ($\mathbf{g'} = -\mathbf{n}$), whereas $\mathbf{v}$ and $\mathbf{n}$ make an acute (actually small) angle $\delta$: $v'_{/\!/} \equiv \mathbf{v'}.\mathbf{n} = \cos\delta$. Then

(6.8) $\qquad\qquad\qquad \mathbf{w} = s_1(\mathbf{v'}_\perp \cos\delta - \mathbf{n}\sin^2\delta).$

(Recall that $s_1$ is of the order of unity.) Thus, $\mathbf{A_S}$ has a component directed towards the Sun, as the residual acceleration found from the data analysis [14], but it has also a larger tangential component: $|\mathbf{v'}_\perp \cos\delta| = \sin\delta \cos\delta \approx \sin\delta \gg \sin^2\delta$ at small $\delta$. Moreover, the maximum magnitude is very large, because $U^{(1)}(\mathbf{a})/|\mathbf{g}| \approx R \approx 20$ au for Pioneer 10 $ca.$ 1980, whence

(6.9) $\qquad\qquad\qquad A_{S\,max} \approx 10^{-3} d$ m/s$^2$ for Pioneer 10.

A reasonable estimate seems to be $\sin\delta \approx 0.1$ (perhaps less). Since $d \approx 1$, the radial component of $\mathbf{A_S}$ would then be $\approx 10^{-5}$ m/s$^2$, this being $10^4$ times larger than the value found from the data analysis [14]. As to the total magnitude of the structure-dependent part of the acceleration, it would then be $|\mathbf{A_S}| \approx 10^{-4}$ m/s$^2$, thus $10^5$ times larger than the residual acceleration [14]!

## 7. CONCLUSION

In any relativistic theory of gravitation, one may expect that the motion of matter be influenced by all forms of energy, hence by kinetic and potential energy and hence also by the distribution of them − that is, by the structure of the bodies. The asymptotic post-Newtonian approximation (PNA) makes this influence appear quite clearly, because it separates the equations of motion corresponding to the successive orders in the expansion with respect to the field-strength parameter. Thus, the equation for the first PN correction to the motion of the mass centers includes several integrals of Newtonian fields, which depend explicitly on the structure. Only in extremely particular cases might this equation be grouped with the Newtonian equation to give a single equation, which would thus include a formula for an ``effective mass" that would assign once for all a particular weight to the different forms of energy. The integrals mentioned [7, Eqs. (4.8-11) for the investigated scalar theory] are defined over the domain occupied by the body itself whose motion is sought for. Since they depend nonlinearly on the matter fields, there is no general reason why the influence of that body itself over its own motion should cancel, even if its size is very small. Precisely, if the PN spatial metric contains space derivatives of the Newtonian potential $U$, then that integral which contains the spatial Christoffel symbols will depend on second space derivatives of $U$. The ``self" part of the latter ones is related to the density and to the geometry of the body in a way which does not depend on its size, hence in that case *the structure of the body itself does influence its own motion even at the point-particle limit.*

In this paper, the limiting acceleration has been determined, for the scalar theory, up to unknown $O(\xi)$ terms, where $\xi$ is the size of the body; complementary checks by the author show that the omitted terms are indeed much smaller. One thus isolates the structure-dependent part $\mathbf{A_S}$ of the acceleration of the small body (in particular, when the external gravitational field is static and spherically symmetric, then the acceleration of the small body differs from the acceleration of a test particle just by $\mathbf{A_S}$). *This is a definite violation of the weak equivalence principle* (WEP) *for **real bodies** (however small they may be), although the validity of the WEP for **test particles** is a built-in feature of the theory, just in the same way as it is in GR. Thus it is a very surprising and consequential result. As such, it needed to be established convincingly – whence the enough-detailed PN calculations.* We have given arguments showing that the presence of a structure-dependent term in the 1PN acceleration might occur also in GR in some gauges if the same approximation scheme were used. *E.g.,* it might occur in a gauge that gives Schwarzschild's standard solution in the spherical static case (a such gauge does exist: see Ref. 23, Note 15). The present results have been summarized in Ref. 24. That WEP violation should not occur in the "relativistic theory of gravitation" of Logunov *et al.* [25], because there, just like in GR under the harmonic gauge, the 1PN spatial metric has the form $\mathbf{g}^{(1)}= (1+2U/c^2)\mathbf{g}^0$ with $\mathbf{g}^0$ an Euclidean metric. The same is true in a modified version [26] of the scalar theory, in which the gravitational space contraction is assumed to be isotropic. Hence, the WEP violation found at the 1PN approximation in the present work, should not occur any more in this modified version. In fact the author has already checked that this violation *does not* occur any more in the new version.

The numerical value of $\mathbf{A_S}$ has been investigated. Although it depends sensitively on orientation and on the precise structure, it reaches high values. On the surface of the Earth, it might conflict with announced results of the tests of the WEP (it would need a detailed investigation to check this precisely). For the Pioneer spacecrafts, it conflicts by some five orders of magnitude with the value extracted from the data analysis for the residual acceleration, i.e. $(8.74\pm 1.33)\times 10^{-10}$ m/s$^2$ [14]. In the author's opinion, the residual acceleration of the Pioneers depends on the theory used to model their motion and to analyse the data, because the equations of motion of the spacecrafts, and of the main celestial bodies as well, are theory-dependent. But the structure-dependent acceleration predicted for the Pioneer spacecrafts by the scalar theory – in its version with anisotropic space contraction – seems just too big. This has been found also by trying to adjust the equations of motion so as to fit the Pioneers' reference trajectories. *Therefore, that version of the scalar theory seems discarded by the present work.* The author is investigating the new version, with isotropic space contraction [26], in which this WEP violation does not take place.

## APPENDIX. FOCK'S "LIAPUNOV'S EQUATION"

In a body (*a*) that is rigidly moving and subjected to Newton's gravitation (in the present case, both assumptions apply to the zero-order PNA), let us define as Fock [1, Eq. (73.16)]

(A1) $$V_a \equiv u_a + \tfrac{1}{2}\Omega_{ik}\Omega_{jk}(x^i - a^i)(x^j - a^j) \equiv u_a + \Omega_a.$$

If other bodies are present, their gravitational field influences the equilibrium of body (*a*), Eq. (3.2). However, if either (i) the bodies are well-separated, so that their mutual distances are all ord($\eta^{-1}$) with $\eta$ a small parameter (see Ref. 10 for a precise asymptotic framework), or (ii) the considered body (*a*) is very small [Eq. (2.1) here], then Eq. (3.2) leads to a relation involving only the self-fields. In case (i), we may write:

(A2) $$U^{(a)}_{,i}(\mathbf{x}) = U^{(a)}_{,i}(\mathbf{a}) + U^{(a)}_{,i,j}(\mathbf{a})(x^j - a^j) + O(\eta^4) \qquad (\mathbf{x} \in D_a),$$

and, using this, Eq. (3.2) gives us {noting that $\dot{\Omega}_{ij} = O(\eta^3)$, as one may easily check from the exact rotational equations valid in Newtonian theory [1, Eq. (72.13)]}:

(A3) $$\rho\,[u_{a,i} + \Omega_{ik}\,\Omega_{jk}\,(x^j - a^j)] = p_{,i} + O(\eta^3).$$

The left-hand side is just $\rho V_{a,i}$, thus

(A4) $$\rho V_{a,i} = p_{,i} + O(\eta^3).$$

In case (ii), we have from (3.1)$_1$, (3.4) and since $x^j - a^j = O(\xi)$:

(A5) $$\rho V_{a,i} = p_{,i} + O(\xi),$$

hence in both cases

(A6) $$\rho V_{a,i} = p_{,i} + O(\delta),$$

with $\delta = \eta^3$ in case (i) and $\delta = \xi$ in case (ii). Consider any segment of straight line, and let $dl = (dx^i dx^i)^{1/2}$ be the Euclidean length element on this segment (in our Cartesian coordinates). Equation (A6) gives us

$$\text{(A7)} \qquad \rho \frac{dV_a}{dl} = \frac{dp}{dl} + O(\delta).$$

On the other hand, the isentropy equation for the barotropic fluid:

$$\text{(A8)} \qquad d\Pi = -p\, d(1/\rho)$$

may be rewritten as

$$\text{(A9)} \qquad \rho \frac{d}{dp}\left(\Pi + \frac{p}{\rho}\right) = 1,$$

and since this is valid pointwise in the barotropic fluid, it implies that, for any segment,

$$\text{(A10)} \qquad \rho \frac{d}{dl}\left(\Pi + \frac{p}{\rho}\right) = \frac{dp}{dl}.$$

Combining (A7) and (A10), and since $\rho = \text{ord}(\delta^0)$ [10, Eq. (3.6), for case (i)], we obtain for any segment

$$\text{(A11)} \qquad \frac{d}{dl}\left(\Pi + \frac{p}{\rho}\right) = \frac{dV_a}{dl} + O(\delta).$$

Integrating this on any segment starting from a fixed point, *e.g.,* from the mass center **a**, we get

$$\text{(A12)} \qquad \Pi + \frac{p}{\rho} = V_a + C + O(\eta^3) \quad \text{inside } D_a$$

in case (i). In case (ii), due to the fact that the length of the segments is itself $O(\xi)$, we get

$$\text{(A13)} \qquad \Pi + \frac{p}{\rho} = V_a + C + O(\xi^2) \quad \text{inside } D_a.$$

(We have assumed that the body is ``star-like"; by considering a chain of segments, the above results can be extended to a connected body.) Since the elastic energy $\Pi$ is defined [by integrating (A8)] only up to a constant, we may choose the latter so that the constant $C$ in (A12), or in (A13), is zero. However, $C$ should depend on time if the body is not stationary, in which case cancelling $C$ at any time would mean that the reference pressure taken to integrate (A8) would depend on time.

Equation (A12) with $C = 0$ is Eq. (73.26) of Fock [1], and it is used to evaluate several integrals that enter in his translational equations of motion. However, the order of the remainder is not given in Ref. 1, moreover the reasoning used to go from (73.19) [Eq. (A4) here] to (73.26) is not fully clear to the present author. Case (ii) (of a small body) was not considered in Ref. 1.

# REFERENCES

1. Fock V., *The theory of space, time and gravitation* (2nd English edn., Pergamon, Oxford), 1964.
2. Chandrasekhar S., *Astrophys. J.,* **142**, 1488 (1965).
3. Futamase T. and Schutz B. F., *Phys. Rev. D,* **28**, 2363 (1983).
4. Rendall A. D., *Proc. Roy. Soc. Lond. A,* **438**, 341 (1992).
5. Arminjon M., *Revue Roumaine Sci. Tech. - Méc. Appl.*, **43**, 135 (1998). [gr-qc/9912041]
6. Arminjon M., *Roman. J. Phys.,* **45**, N° 5-6, 389 (2000). [gr-qc/0003066]
7. Arminjon M., *Roman. J. Phys.,* **45**, N° 9-10, 645 (2000). [astro-ph/0006093 §§ 1-4]
8. Arminjon M., a) *Roman. J. Phys.,* **45**, N° 9-10, 659 (2000).
   A more detailed version is: b) astro-ph/0006093 §§ 5-7 and appendices.
9. Arminjon M., *Nuovo Cimento B,* **116**, 1277 (2001). [gr-qc/0106087]
10. Arminjon M., *Roman. J. Phys.,* to appear (2003/2004). [gr-qc/0202029]
11. Arminjon M., *Theor. Math. Phys.,* **140,** N°1, 1011 (2004) [*Teor. Mat. Fiz.,* **140,** N°1, 139 (2004)]. [gr-qc/0310062]
12. Arminjon M., in *Recent Research Developments in Astronomy & Astrophyics.* (S. G. Pandalai, ed., Research Sign Post, Trivandrum), Vol. 1 (2003), pp. 859-879. [gr-qc/0305078]
13. Anderson J. D., Laing P. A., Lau E. L., Liu A. S., Nieto M. M. and Turyshev S., *Phys. Rev. Lett.,* **81**, 2858 (1998). [gr-qc/9808081]
14. Anderson J. D., Laing P. A., Lau E. L., Liu A. S., Nieto M. M. and Turyshev S., *Phys. Rev. D,* **65**, 082004 (2002). [gr-qc/0104064]
15. Arminjon M., in *Fifth Conf. "Physical Interpretations of Relativity Theory", Supplementary Papers* (M. C. Duffy, ed., University of Sunderland/ Brit. Soc. Phil. Sci., 1998), pp. 1-27. [Online at http://geo.hmg.inpg.fr/arminjon/PIR96_4B.htm]
16. Weinberg S., *Gravitation and Cosmology* (J. Wiley & Sons, New York, 1972).
17. Standish E.M., *Astron. & Astrophys.,* **101**, L17 (1981).
18. Bretagnon P., *Astron. & Astrophys.,* **114**, 278 (1982).
19. Newhall X. X., Standish E.M. and Williams J.G., *Astron. & Astrophys.,* **125**, 150 (1983).
20. Arminjon M., *Int. J. Mod. Phys. A,* **17**, 4203 (2002). [gr-qc/0205105]
21. Standish E.M., Newhall X. X., Williams J.G. and Folkner W. F., Jet Prop. Lab. Interoffice Memo. 314.10-127 (1995).
22. Fischbach E. and Talmadge C. L., *The Search for Non-Newtonian Gravity* (Springer, New York-Berlin-Heidelberg, 1998), chapter 4.
23. Arminjon M., in *Relativistic World Ether* (M. C. Duffy & L. Kostro, eds., to be published). [gr-qc/0401021]
24. Arminjon M., in *Gravitational Waves and Experimental Gravity, Proc. 38th Rencontres de Moriond* (J. Dumarchez & J. Tran Thanh Van, eds., The Gioi, Hanoi, 2004), pp. 377-382. [gr-qc/0306025]
25. Logunov, A.A. and Mestvirishvili, M.A., *The Relativistic Theory of Gravitation* (Mir, Moscow, 1989).
26. Arminjon M., in *Ninth Conf. "Physical Interpretations of Relativity Theory", Proceedings* (M. C. Duffy, ed., to appear). [physics/0404132]